\documentclass{article}
\PassOptionsToPackage{numbers}{natbib}
\usepackage[preprint]{neurips_2024}

\usepackage[utf8]{inputenc}
\usepackage[T1]{fontenc}
\usepackage{hyperref}
\usepackage{url}
\usepackage{booktabs}
\usepackage{amsmath}
\usepackage{amssymb}
\usepackage{graphicx}
\usepackage{multirow}
\usepackage{tabularx}
\usepackage{array}
\usepackage{microtype}
\usepackage{xcolor}
\usepackage{float}
\usepackage{caption}
\usepackage{subcaption}
\usepackage{enumitem}
\usepackage{placeins}

\usepackage{tikz}
\usetikzlibrary{shapes.geometric, arrows.meta, positioning}

\title{LLM Inference at the Edge: Mobile, NPU, and GPU\\Performance Efficiency Trade-offs Under Sustained Load}

\author{}

\tikzset{
  startstop/.style={rectangle, rounded corners=8pt, minimum width=3.5cm,
    minimum height=0.7cm, text centered, draw=black, fill=gray!20, font=\small},
  process/.style={rectangle, minimum width=3.5cm, minimum height=0.7cm,
    text centered, draw=black, fill=blue!10, font=\small},
  decision/.style={diamond, aspect=2.5, minimum width=3.5cm, minimum height=0.7cm,
    text centered, draw=black, fill=orange!15, font=\small},
  arrow/.style={-{Stealth[length=2mm]}, thick},
}

\usepackage{listings}
\begin{document}

\maketitle
% Then after \maketitle, add:
\vspace{-2em}
\begin{center}
\begin{tabular}{>{\centering\arraybackslash}p{0.45\textwidth}>{\centering\arraybackslash}p{0.45\textwidth}}
  \textbf{Pranay Tummalapalli} &
  \textbf{Sahil Arayakandy} \\[3pt]
  Conscious Engines &
  Conscious Engines \\[3pt]
  \texttt{\small pranay@consciousengines.com} &
  \texttt{\small sahil@consciousengines.com} \\[12pt]
  \textbf{Ritam Pal} &
  \textbf{Kautuk Kundan} \\[3pt]
  Conscious Engines &
  Conscious Engines \\[3pt]
  \texttt{\small ritam@consciousengines.com} &
  \texttt{\small kautuk@consciousengines.com} \\
\end{tabular}
\end{center}
\vspace{1em}

% ============================================================
\begin{abstract}
% ============================================================

Deploying large language models on-device for always-on personal agents demands
sustained inference from hardware tightly constrained in power, thermal envelope,
and memory. We benchmark Qwen~2.5~1.5B (4-bit quantised) across four platforms:
a Raspberry Pi~5 with Hailo-10H NPU, a Samsung Galaxy S24 Ultra, an iPhone~16
Pro, and a laptop NVIDIA RTX~4050 GPU. Using a fixed 258-token prompt over 20
warm-condition iterations per device under greedy decoding, we measure
throughput, latency, power, and thermal behaviour. For mobile platforms,
thermal management supersedes peak compute as the primary constraint: the
iPhone~16 Pro loses roughly 40\% of its peak throughput within three
iterations and settles into a Hot-state plateau near 23.7~tok/s, while the
S24 Ultra degrades $\sim$15\% over 20 iterations into a Snapdragon
governor-regulated plateau near 10~tok/s.
On dedicated hardware, distinct constraints dominate: the RTX~4050 is bounded
by its battery power ceiling, while the Hailo-10H is limited by on-module
memory bandwidth. The RTX~4050 sustains 131.7~tok/s at 34.1~W; the Hailo-10H
sustains 6.9~tok/s at under 2~W with near-zero variance, matching the RTX~4050
in energy proportionality at $19\times$ lower throughput. Results should be
interpreted as platform-level deployment characterisations for a single model
and prompt type, reflecting hardware and software combined, rather than
general claims about hardware capability alone.
\end{abstract}

% ============================================================
\section{Introduction}
% ============================================================

The emergence of sub-2B parameter language models capable of useful instruction
following~\cite{qwen25,touvron2023llama2} has renewed interest in on-device
LLM deployment. Running inference locally offers offline capability, reduced latency, and privacy, properties that are particularly valuable for
always-on personal agents that must respond to continuous streams of user
queries without round-tripping to a cloud server.

Modern mobile SoCs include dedicated neural processing units with tens of TOPS
of claimed throughput~\cite{qualcomm2023,apple2024coreml}, and quantisation
techniques such as GPTQ~\cite{frantar2023gptq} and AWQ~\cite{lin2023awq}
compress models to under 1~GB, making them memory-feasible on phones and
single-board computers. Yet the gap between peak hardware specifications and
\emph{sustained} real-world inference performance under thermal load remains
poorly characterised. Existing benchmarks either target single-inference
latency\cite{ignatov2018,primate2024}, cloud-hosted
throughput~\cite{reddi2020mlperf}, or report thermal effects only as a
side observation rather than a primary focus~\cite{laskaridis2024melting}.
The MELTing Point framework~\cite{laskaridis2024melting} provides the closest
prior work, characterising LLM inference on mobile devices with attention to
energy and thermal behaviour, but does not run extended back-to-back query
sequences sufficient to reveal the full thermal degradation curve.
Xiao et al.~\cite{xiao2024llmpockets} benchmark LLMs on mobile platforms with attention
to throughput and observe 10--20\% degradation over five consecutive runs,
but do not characterise steady-state thermal floors or test beyond a handful
of iterations. MLPerf Inference~\cite{reddi2020mlperf} now includes edge LLM
scenarios as of v5.1, but measures peak performance without tracking thermal
throttling or sustained-workload behaviour.

On the hardware side, quantised LLM inference on dedicated edge NPUs remains
largely unexplored in the literature. To the best of our knowledge, no independent benchmark of LLM inference
on Hailo hardware has been published; available figures derive from
vendor publications alone.

This leaves three questions unanswered: how mobile inference performance
degrades over repeated consecutive queries; whether low-power NPU
accelerators can match mobile devices in sustained throughput per watt;
and what the practical trade-offs are between a dedicated GPU, a mobile SoC,
and an edge NPU for always-on agent deployments.

To answer these, we conduct controlled warm-condition benchmarking of
Qwen~2.5~1.5B across four platforms, a Raspberry Pi~5 with Hailo-10H NPU,
a Samsung Galaxy S24 Ultra, an iPhone~16 Pro, and a laptop with NVIDIA
RTX~4050 GPU, using a standardised prompt and uniform metric collection
across 20 iterations per device. Our principal contributions are: an
independent cross-platform benchmark of sustained LLM inference including a
dedicated edge NPU; empirical characterisation of thermal degradation curves
on two flagship smartphones showing qualitatively different failure modes;
and evidence that the Hailo-10H NPU achieves thermally stable, near-zero-variance
inference at energy proportionality comparable to a laptop GPU, positioning
dedicated edge NPUs as a compelling platform for always-on, power-sensitive
deployments. However, the Hailo-10H's current throughput of 6.9~tok/s, constrained
single-sequence autoregressive decode characteristics, represents a significant
latency limitation for interactive applications, and should be interpreted as
a platform deployment ceiling rather than a fundamental bound on NPU-class
inference.

% ============================================================
\section{Background}
% ============================================================

\subsection{Quantised Inference on Edge Hardware}

Post-training quantisation reduces weight precision from BF16/FP16 to INT4,
yielding approximately $4\times$ model size reduction with minimal perplexity
degradation on instruction-tuned models~\cite{frantar2023gptq,lin2023awq}.
Q4\_0 (4-bit grouped quantisation) has become the dominant format for on-device
deployment due to its compatibility with GGUF, MLC-LLM, and MLX runtimes,
while GPTQ~\cite{frantar2023gptq} provides an alternative calibrated
quantisation path used by vLLM and PyTorch-based serving stacks.
llama.cpp~\cite{gerganov2023llamacpp} provides cross-platform GGUF-format
inference across targets.

\subsection{Autoregressive Decoding Phases}

LLM inference comprises two phases with distinct compute characteristics.
During \emph{prefill}, the input prompt is processed in parallel; this phase
is compute-bound, proportional to prompt length, and dominates
time-to-first-token (TTFT). During \emph{decode}, tokens are generated one
at a time autoregressively; this phase is memory-bandwidth-bound, with
throughput determined by KV-cache access speed. Peak hardware TOPS ratings reflect throughput under compute-bound,
high-utilisation conditions. Prefill is relatively compute-bound and
scales with available TOPS; sustained decode throughput is
memory-bandwidth-bound, with throughput determined by KV-cache access
speed, a key distinction when evaluating always-on agents that generate
long responses.

\subsection{Thermal Throttling in Mobile SoCs}

Mobile SoCs rely on passive cooling and must respect strict power budgets
to avoid skin temperature violations. When junction temperatures exceed
thermal thresholds, Dynamic Voltage and Frequency Scaling (DVFS) reduces
clock frequencies, directly reducing throughput. Sustained LLM inference
is particularly susceptible to this mechanism. Recent work on mobile DVFS
governors~\cite{zhang2025dvfs} has shown that Energy-Aware Scheduling can
cause CPUs to run at suboptimally low frequencies during LLM inference,
degrading latency and energy efficiency, illustrating that OS-level power
management interacts with inference workloads in ways beyond simple thermal
capping.

% ============================================================
\section{Methodology}
% ============================================================
We select four platforms spanning the performance--efficiency spectrum relevant
to personal agent deployment: a Raspberry Pi~5 paired with a Hailo-10H M.2
NPU module, representing always-on low-power edge deployment; a Samsung
Galaxy S24 Ultra as a flagship Android device with Snapdragon 8 Gen 3; an
iPhone~16 Pro as a flagship iOS device with Apple A18 Pro; and a laptop with
NVIDIA RTX~4050 GPU as a battery-powered edge device in a laptop form factor.
Full hardware and software specifications are given in Tables~\ref{tab:hardware}
and~\ref{tab:software}.

\subsection{Platforms and Inference Stacks}

The Hailo-10H connects via PCIe Gen~3.0, providing 40~TOPS at under 5~W.
hailo-ollama partitions model layers between NPU and CPU, with attention
offloaded to the NPU and Q4 weights dispatched to the NPU's INT8 compute
units via the Hailo Dataflow Compiler. Power is sampled from the RPi~5
PMIC rails via an INA219 at 1~kHz; this reflects total system draw rather
than isolated NPU power, as no dedicated current sensor sits on the
Hailo-10H's PCIe supply.

The Samsung Galaxy S24 Ultra runs the model compiled to TVM binary format
targeting the Adreno~750 GPU via Apache TVM 0.14 through MLC-LLM, as a
headless Android service on Android 16 with the display held off during
the measured run. The runtime uses a 2{,}048-token context and a 128-token
prefill chunk; the reduced chunk inflates prefill latency relative to
Adreno hardware capability (see Section~\ref{sec:discussion}). Battery
current, voltage, and temperature are sampled from the Android
\texttt{BatteryManager} and \texttt{HealthService} APIs per iteration, with
absolute accuracy caveated.

The iPhone~16 Pro runs a custom SwiftUI app integrating MLX Swift
(v0.29.1), with inference dispatched to the A18 Pro GPU via Metal; MLX
does not target the Neural Engine. Greedy decoding
(\texttt{temperature = 0.0}) with \texttt{maxKVSize = 2048} fixes output
length and KV footprint across iterations. iOS exposes no per-component
power draw to third-party apps, so thermal state
(\texttt{ProcessInfo.thermalState}) and battery level (\texttt{UIDevice})
are sampled per iteration as the sole energy proxy.

The RTX~4050 laptop serves the model via vLLM with the PyTorch backend on
CUDA 12.1; \texttt{nvidia-smi} logs GPU power, utilisation, and temperature
at 100~ms. Runs were conducted on battery, which imposes a sustained power
ceiling near 35~W (${\sim}45\%$ of the 75~W TGP), so reported throughput
is battery-throttled.

\begin{table}[!htbp]
\centering
\caption{Hardware specifications across evaluated platforms.}
\label{tab:hardware}
\small
\begin{tabularx}{\textwidth}{lXXXX}
\toprule
 & \textbf{RPi~5 + Hailo-10H} & \textbf{S24 Ultra} & \textbf{iPhone~16 Pro} & \textbf{RTX~4050 Laptop} \\
\midrule
SoC / CPU     & BCM2712 4$\times$A76 @ 2.4~GHz & Snapdragon 8 Gen 3 & Apple A18 Pro & Intel Core i5-13420H \\
AI Accel.     & Hailo-10H NPU (40~TOPS, $<$5~W) & Hexagon NPU (45~TOPS) & Neural Engine 17-core ($\sim$35~TOPS) & RTX~4050 (80 Tensor Cores) \\
RAM           & 8~GB LPDDR4X & 12~GB LPDDR5X & 8~GB Unified & 16~GB DDR5 \\
GPU           & VideoCore VII & Adreno 750 & 6-core Apple GPU & RTX~4050 (2560 CUDA cores) \\
VRAM          & Shared & Shared & Shared & 6~GB GDDR6 \\
OS            & Raspberry Pi OS & Android 16 & iOS 26 & Ubuntu 22.04.3 \\
Idle power    & $\sim$3.5~W & $\sim$0.8~W & $\sim$0.6~W & $\sim$15~W$^\S$ \\
Max power     & $\sim$12~W & $\sim$12~W & $\sim$10~W & 75~W TGP$^\dagger$ \\
\bottomrule
\end{tabularx}
\end{table}

\noindent$^\dagger$Acer Nitro V RTX~4050 laptop GPU TGP is 75~W (60~W base
+ 15~W Dynamic Boost). Benchmarks conducted on battery; sustained inference
power was observed at $\sim$34~W.

$^\S$System idle power $\sim$15~W; GPU idle power $\sim$2~W per \texttt{nvidia-smi}.

\begin{table}[!htbp]
\centering
\caption{Software stack versions across platforms.}
\label{tab:software}
\small
\begin{tabularx}{\textwidth}{lXXXX}
\toprule
 & \textbf{RPi~5 + Hailo-10H} & \textbf{S24 Ultra} & \textbf{iPhone~16 Pro} & \textbf{RTX~4050 Laptop} \\
\midrule
Framework      & hailo-ollama       & MLC-LLM 0.1.0        & MLX Swift v0.29.1      & vLLM (CUDA 12.1)     \\
Backend        & HailoRT 4.17.0     & OpenCL 3.0 + TVM 0.14 & Metal (MLX)           & PyTorch 2.1.0        \\
Language       & Python 3.10        & Java / NDK r26b       & Swift 5.9 (SwiftUI)   & Python 3.10          \\
Model format   & GGUF Q4\_0         & MLC binary q4f16\_2   & MLX safetensors Q4\_0 & GPTQ Int4 safetensors \\
Context window$^\ddagger$ & 2,048             & 2,048             & 2,048 (KV)        & 2,048     \\
Power monitor  & INA219 (1~kHz)     & \texttt{BatteryManager} (per-iter)$^*$ & \texttt{UIDevice} batteryLevel & \texttt{nvidia-smi} (100~ms) \\
\bottomrule
\end{tabularx}

\vspace{0.3em}
\begin{minipage}{\textwidth}
\footnotesize
\noindent$^\ddagger$Context window was fixed to 2{,}048 tokens on all four
platforms to hold the KV-cache memory footprint constant across the
comparison, versus the model's native 32{,}768-token window. The runtime
knob differs per stack: \texttt{context\_window\_size} on MLC-LLM (S24
Ultra), \texttt{maxKVSize} on MLX (iPhone~16 Pro), \texttt{num\_ctx} on
Ollama (RPi~5 + Hailo-10H), and \texttt{max\_model\_len} on vLLM
(RTX~4050).

\noindent$^*$Android \texttt{BatteryManager} readings are collected with the
display held off throughout the measured run; absolute values are treated
with caution (see Section~\ref{sec:discussion}).
\end{minipage}
\end{table}

\FloatBarrier

We select Qwen~2.5~1.5B~\cite{qwen25} with 4-bit quantisation as the
benchmark model. Selection criteria were threefold: native support across all
four inference frameworks, a sub-1~GB memory footprint fitting all device
constraints, and a uniform quantisation level across platforms to minimise
compression as a confounding variable. Each platform uses a native format
conversion of the same base weights: GGUF (hailo-ollama), TVM binary
(MLC-LLM), MLX safetensors (iPhone~16 Pro), and HuggingFace safetensors
(vLLM). Key specifications are listed in Table~\ref{tab:model}.

\begin{table}[!htbp]
\centering
\caption{Qwen~2.5~1.5B model specifications.}
\label{tab:model}
\small
\begin{tabular}{ll}
\toprule
\textbf{Attribute} & \textbf{Value} \\
\midrule
Parameters          & 1.5B \\
Architecture        & Transformer decoder, GQA (2 groups) \\
Layers              & 28 \\
Hidden size         & 1536 \\
Attention heads     & 12 \\
Vocabulary          & 151,936 tokens \\
Context window      & 32,768 tokens \\
\bottomrule
\end{tabular}
\end{table}

\FloatBarrier

\subsection{Prompt and Generation Configuration}

All platforms receive an identical prompt (UTF-8, 258 tokens as tokenised
by the Qwen~2.5 BPE tokenizer) that elicits long-form structured output,
imposing sustained decode load to stress thermal management and
memory-bandwidth utilisation rather than burst performance. Effective
prefill length including framework-applied chat templates is approximately
270--280 tokens and may vary marginally across inference stacks.

\begin{lstlisting}
Write an in-depth, structured, and self-contained essay explaining the
concept of consciousness from multiple perspectives. Begin by clearly
defining consciousness and why it is a difficult concept to study. Then
explore the topic from the following viewpoints, dedicating multiple
detailed paragraphs to each:

  1. Philosophical perspectives, including classical and modern views,
     major debates, and unresolved questions.
  2. Neuroscientific perspectives, covering brain structures, neural
     correlates, current theories, and experimental approaches.
  3. Cognitive science and psychology perspectives, including perception,
     attention, self-awareness, and consciousness disorders.
  4. Artificial intelligence and machine consciousness, discussing whether
     machines can be conscious, functional vs. phenomenal consciousness,
     and current limitations of AI systems.
  5. Ethical and societal implications of understanding or engineering
     consciousness.

Throughout the essay: use clear section headings; explain all technical
terms in plain language; provide illustrative examples where helpful;
compare and contrast different viewpoints; avoid bullet lists unless
necessary for clarity. Conclude with a reflective summary discussing what
remains unknown and what future research directions may look like. The
response should be detailed, continuous, and written in a neutral,
academic tone.
\end{lstlisting}

All platforms decode greedily (temperature~$=0$, top-$k=1$) until the
model emits EOS, so that output length and decode workload are bit-identical
across the 20 iterations on a given platform. Under this setting the natural
stopping length is 646 tokens on the S24 Ultra (MLC-LLM), 819 tokens on the
iPhone~16 Pro (MLX), 564 tokens on the RPi~5 + Hailo-10H, and 1{,}789 tokens
on the RTX~4050 (vLLM). The cross-platform variation reflects differences in
tokenisers and chat-template formatting rather than a configurable cap.

\subsection{Metrics}

Table~\ref{tab:metrics} defines all metrics collected during benchmarking.
Not all metrics are available on all platforms: thermal state is iOS-only,
GPU frequency is Android-only, and power and energy-per-token are reported
only for the RTX~4050 and Hailo-10H due to measurement limitations on mobile
platforms discussed in Section~\ref{sec:discussion}.

\begin{table}[htbp]
\centering
\caption{Metrics collected across platforms.}
\label{tab:metrics}
\small
\begin{tabular}{llp{7cm}}
\toprule
\textbf{Metric} & \textbf{Unit} & \textbf{Definition} \\
\midrule
Decode tokens     & count   & Tokens generated during the decode phase \\
Decode time       & ms      & Wall-clock duration of the decode phase \\
Prefill Time      & ms      & Time from prompt submission to first output token; equivalent to the duration of the prompt encoding phase \\
Throughput (TPS)  & tok/s   & $N_{\text{decode}} / t_{\text{decode}}$ \\
Avg power         & W       & Mean system power draw over inference duration \\
Peak power        & W       & Maximum instantaneous power observed \\
Energy/token      & mJ      & $P_{\text{avg}} \cdot t_{\text{decode}} / N_{\text{decode}}$ \\
CPU / GPU temp    & °C      & Maximum temperature recorded during inference \\
Thermal state     & ---     & iOS categorical signal: Normal, Warm, or Hot \\
Battery drain     & \% SoC  & Change in state of charge over full benchmark run \\
GPU frequency     & MHz     & Observed GPU clock frequency per iteration (Android only) \\
\bottomrule
\end{tabular}
\end{table}

\FloatBarrier

\subsection{Experimental Protocol}

We focus on warm-condition inference: the model is loaded once and remains
in memory across all iterations, representing realistic interactive assistant
usage where queries arrive consecutively. Cold-start model loading time is
recorded where observable but excluded from throughput and latency analysis.

Before each benchmark session, the device is allowed to equilibrate for
10 minutes at ambient temperature (22°C $\pm$ 2°C). After loading the model
and discarding one warm-up inference, we verify thermal stability
($\Delta T < 2$°C over 60 seconds) before executing 20 iterations with a
1-second inter-iteration gap. Per-iteration results are exported to CSV and
validated for token count anomalies.

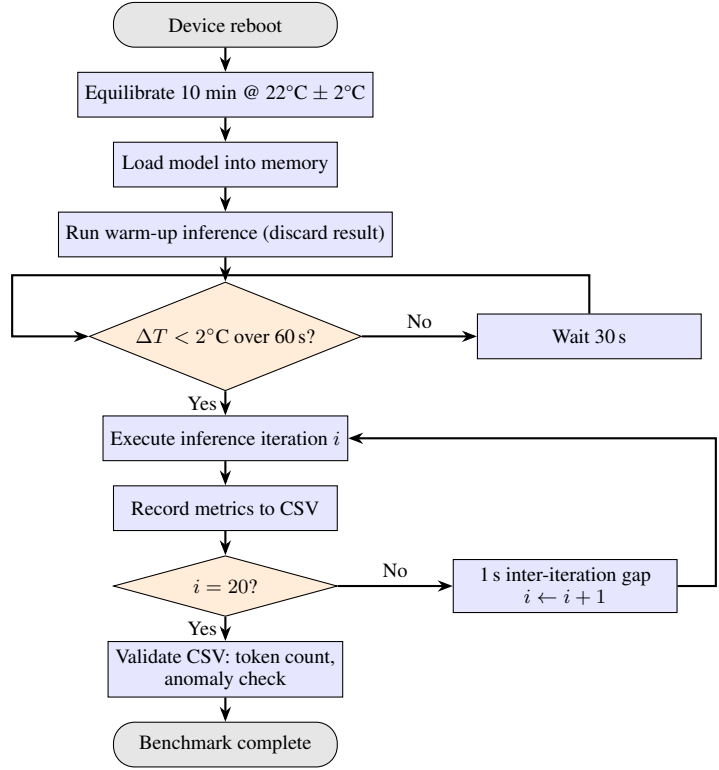
\begin{figure}[H]
\centering
\begin{tikzpicture}[node distance=0.38cm, scale=0.85, every node/.style={transform shape}]
  \node (start)   [startstop]                          {Device reboot};
  \node (equil)   [process, below=of start]            {Equilibrate 10 min @ 22°C $\pm$ 2°C};
  \node (load)    [process, below=of equil]            {Load model into memory};
  \node (warmup)  [process, below=of load]             {Run warm-up inference (discard result)};
  \node (stable)  [decision, below=of warmup]          {$\Delta T < 2^\circ$C over 60\,s?};
  \node (wait)    [process, right=1.8cm of stable]     {Wait 30\,s};
  \node (iter)    [process, below=of stable]           {Execute inference iteration $i$};
  \node (csv)     [process, below=of iter]             {Record metrics to CSV};
  \node (done)    [decision, below=of csv]             {$i = 20$?};
  \node (gap) [process, align=center, right=1.8cm of done] {1\,s inter-iteration gap\\ $i \leftarrow i + 1$};

  \node (valid) [process, align=center, below=of done]        {Validate CSV: token count,\\anomaly check};
  \node (stop)    [startstop, below=of valid]          {Benchmark complete};

  \draw[arrow] (start)   -- (equil);
  \draw[arrow] (equil)   -- (load);
  \draw[arrow] (load)    -- (warmup);
  \draw[arrow] (warmup)  -- (stable);
  \draw[arrow] (stable)  -- node[anchor=east, font=\small]{Yes} (iter);
  \draw[arrow] (stable)  -- node[anchor=south, font=\small]{No} (wait);
  \draw[arrow] (wait.north) -- ++(0,0.6) -| ([xshift=-1.2cm]stable.west) -- (stable.west);
  \draw[arrow] (iter)    -- (csv);
  \draw[arrow] (csv)     -- (done);
  \draw[arrow] (done)    -- node[anchor=east, font=\small]{Yes} (valid);
  \draw[arrow] (done)    -- node[anchor=south, font=\small]{No} (gap);
  \draw[arrow] (gap.east) -- ++(0.6,0) |- (iter.east);
  \draw[arrow] (valid)   -- (stop);
\end{tikzpicture}
\caption{Experimental protocol flowchart. Each platform undergoes this
         procedure independently. Cold-start prefill-time from the warm-up
         inference is recorded separately and excluded from throughput analysis.}
\label{fig:protocol}
\end{figure}

% \FloatBarrier
\vspace{-0.5em}

% ============================================================
\section{Results}
% ============================================================

\subsection{NVIDIA RTX~4050 (Desktop GPU)}

The RTX~4050 provides the performance baseline in this study, representative
of a battery-powered edge device in a laptop form factor.
Table~\ref{tab:rtx_summary} summarises results across 20 warm iterations;
run~1 includes cold-start overhead and is excluded from steady-state
statistics.

\begin{table}[H]
\centering
\caption{RTX~4050 inference performance (runs 2--20, warm condition).
         Power values derived from system-level measurement at 100\,ms polling.}
\label{tab:rtx_summary}
\small
\begin{tabular}{lrrrr}
\toprule
\textbf{Metric} & \textbf{Mean} & \textbf{Std Dev} & \textbf{Min} & \textbf{Max} \\
\midrule
Decode tokens            & 1789    & 0      & 1789   & 1789   \\
Decode time (ms)         & 13{,}697 & 274   & 12{,}881 & 13{,}906 \\
Throughput (tok/s)       & 131.70  & 2.87   & 128.65 & 138.88 \\
Prefill time (ms)        & 1{,}998  & 42    & 1{,}894 & 2{,}028 \\
Avg system power (W)     & 34.12   & 0.17   & 33.75  & 34.40  \\
Peak system power (W)    & 35.28   & 0.21   & 35.05  & 35.73  \\
Energy per token (mJ)    & 297.3   & 6.5    & 282.7  & 304.9  \\
GPU temperature (°C)     & 65.4    & 4.3    & 55     & 70     \\
CPU temperature (°C)     & 87.5    & 4.2    & 77     & 93     \\
Battery drain            & \multicolumn{4}{l}{12\% over 20 runs (100\% $\to$ 88\%)} \\
\bottomrule
\end{tabular}
\end{table}

\begin{figure}[H]
\centering
\includegraphics[width=\textwidth]{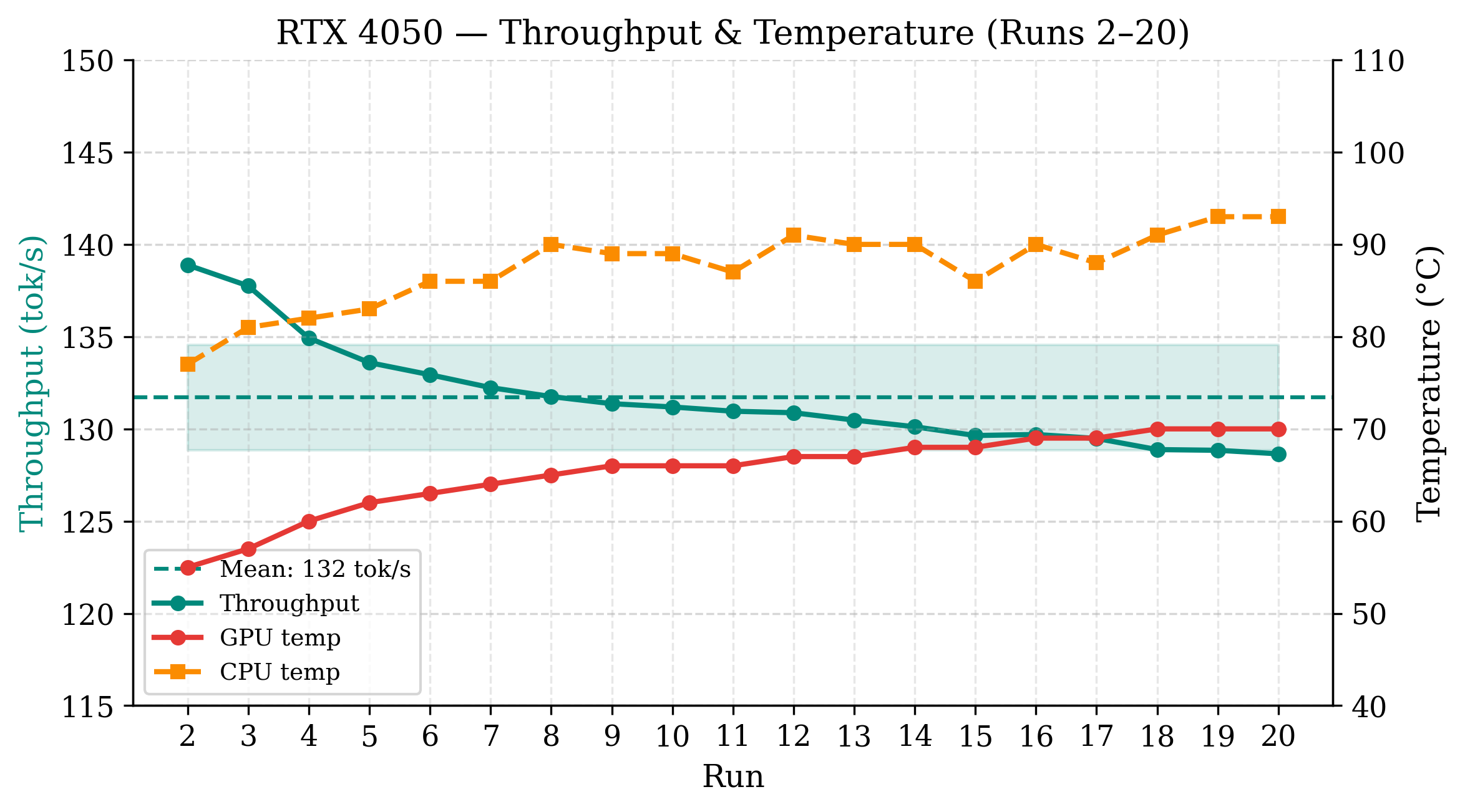}
\caption{RTX~4050 per-run throughput (left axis) and GPU/CPU temperature
         (right axis) across runs 2--20. Mean throughput of 131.70~tok/s
         (CV~= 2.2\%) confirms stable battery-throttled performance;
         GPU rises from 55°C to 70°C with no throttling observed.}
\label{fig:rtx_combined}
\end{figure}

The RTX~4050 sustains a mean throughput of 131.7~tok/s ($\sigma = 2.87$,
CV~= 2.2\%) with deterministic output length (1789 tokens per run). Run~1 is
excluded as it reflects cold-start model loading (93.8~tok/s, 52.8~W average
power). Under battery power, average power stabilises at $34.1 \pm 0.2$~W
with transient peaks reaching 35.73~W, confirming the workload is
memory-bandwidth bound rather than compute bound. GPU temperature rises
from 55°C at run~2 to 70°C by run~20, reflecting gradual heat buildup,
though no thermal throttling is observed---the throughput CV of 2.2\%
confirms stable performance throughout. Energy per token stabilises at
$297 \pm 7$~mJ across warm runs.

\FloatBarrier

\subsection{Raspberry Pi~5 + Hailo-10H NPU}

Table~\ref{tab:rpi_results} summarises results across 20 runs. Run~1
includes a cold-start prefill-time of 11.86~s reflecting initial model loading
into NPU memory and is excluded from steady-state statistics.

\begin{table}[H]
\centering
\caption{RPi~5 + Hailo-10H inference performance (runs 2--20, warm condition).}
\label{tab:rpi_results}
\begin{tabular}{lrrrr}
\toprule
\textbf{Metric} & \textbf{Mean} & \textbf{Std Dev} & \textbf{Min} & \textbf{Max} \\
\midrule
Decode tokens              & 564     & 0      & 564     & 564     \\
Decode time (ms)           & 81{,}569 & 34    & 81{,}516 & 81{,}640 \\
Throughput (tok/s)         & 6.914   & 0.003  & 6.908   & 6.919   \\
Prefill time (ms)          & 1{,}287  & 1     & 1{,}286  & 1{,}290 \\
Avg system power (W)       & 1.870   & 0.016  & 1.851   & 1.921   \\
Peak system power (W)      & 3.065   & 1.099  & 2.383   & 6.662   \\
Energy per token (mJ)      & 270.5   & 2.4    & 267.6   & 277.9   \\
CPU temperature (°C)       & 52.7    & 0.8    & 51.0    & 53.8    \\
NPU temperature (°C)       & 58.5    & 1.1    & 55.9    & 60.0    \\
\bottomrule
\end{tabular}
\end{table}

\begin{figure}[H]
\centering
\includegraphics[width=0.85\textwidth]{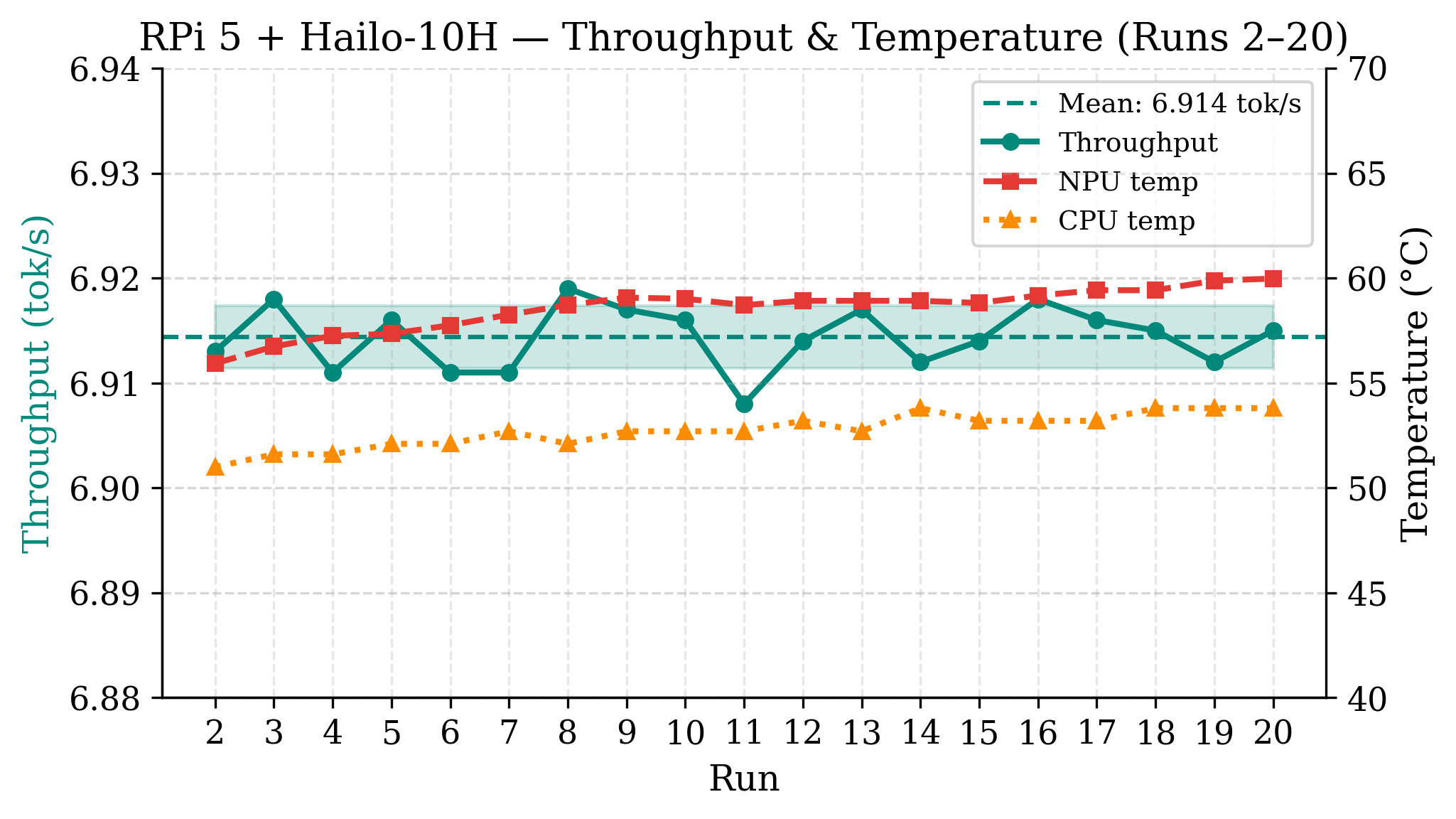}
\caption{RPi~5 + Hailo-10H throughput (line, left axis) and NPU/CPU
         temperature (lines, right axis) across runs 2--20.
         Throughput CV of 0.04\% and stable temperatures confirm
         no throttling at any point.}
\label{fig:hailo_tps_temp}
\end{figure}

The Hailo-10H delivers the most consistent inference of any platform in this
study. With greedy decoding, output length is fully deterministic, all 19
warm runs produce exactly 564 tokens, yielding a decode time CV of 0.04\%
and a throughput CV of 0.04\%, effectively zero variance. Prefill is similarly
invariant at $1{,}287 \pm 1$~ms (CV~= 0.11\%), reflecting the deterministic
dispatch behaviour of the NPU hardware. Average system power (RPi~5 +
Hailo-10H combined) is $1.87 \pm 0.02$~W, with energy per token of
$270.5 \pm 2.4$~mJ, the lowest of any platform in this study. Thermal behaviour
is exemplary: CPU temperature is confined to $52.7 \pm 0.8$°C and NPU
temperature to $58.5 \pm 1.1$°C, with no upward trend and no throttling
observed at any point across all 20 iterations.

The observed throughput of 6.914 tok/s is substantially below the Hailo-10H's rated 40 TOPS peak compute figure, which is expected given the nature of autoregressive LLM decode. Vendor TOPS ratings reflect peak throughput under idealised batch workloads; sustained single-sequence decode is memory-bandwidth bound rather than compute bound, and cannot exploit the NPU's parallel compute units efficiently. The result is broadly consistent with Hailo's own published figure of 6.82 tok/s for Qwen2.5-1.5B-Instruct on the same hardware configuration, suggesting the deployment is operating near the practical ceiling for this platform. The primary bottleneck is on-module LPDDR4 memory bandwidth rather than PCIe: although the Hailo-10H exposes a PCIe Gen 3 ×4 interface (~4 GB/s theoretical), the Raspberry Pi 5 provides only a PCIe Gen 2 ×1 link (~500 MB/s theoretical, ~400 MB/s effective), this constraint does not appear to be the binding limiter given that our measured throughput matches Hailo's reference figure obtained under identical link conditions. Additional deployment-specific constraints include CPU--NPU layer partitioning in hailo-ollama, where attention layers not offloaded to the NPU execute on the ARM Cortex-A76, and per-token HailoRT dispatch overhead over PCIe. The 6.914 tok/s result should therefore be interpreted as a memory-bandwidth-bound deployment figure reflecting the practical ceiling of this platform, rather than a reflection of the NPU's peak compute capability.

\FloatBarrier

\subsection{iPhone~16 Pro (iOS / MLX)}

Table~\ref{tab:iphone_results} reports available metrics across 20 iterations
with greedy decoding and a bounded 2{,}048-token KV cache. Generation length
is deterministic at 819 tokens per iteration, so all variance reflects
scheduler and thermal effects rather than sampling noise. Per-inference
power draw and energy-per-token are not reported for this platform due to
iOS measurement limitations.

\begin{table}[H]
\centering
\caption{iPhone~16 Pro inference performance (20 iterations, warm condition,
         greedy decoding, maxKVSize = 2{,}048).}
\label{tab:iphone_results}
\begin{tabular}{lrrrr}
\toprule
\textbf{Metric} & \textbf{Mean} & \textbf{Std Dev} & \textbf{Min} & \textbf{Max} \\
\midrule
Decode tokens              & 819     & 0      & 819    & 819    \\
Decode time (s)            & 32.19   & 3.69   & 20.23  & 35.06  \\
Throughput (tok/s)  & 25.89 & 4.11  & 23.36 & 40.49 \\
TTFT (s)            & 0.466 & 0.064 & 0.268 & 0.544 \\
Prompt throughput (tok/s)  & 606.5   & 122.9  & 505.7  & 1027.5 \\
Battery drain (total)      & \multicolumn{4}{l}{5\% over 20 iterations (100\% $\to$ 95\%)} \\
Thermal state              & \multicolumn{4}{l}{Normal (iter 1--2), Warm (3--16), Hot (17--20)} \\
\bottomrule
\end{tabular}
\end{table}

\begin{figure}[H]
\centering
\includegraphics[width=0.85\textwidth]{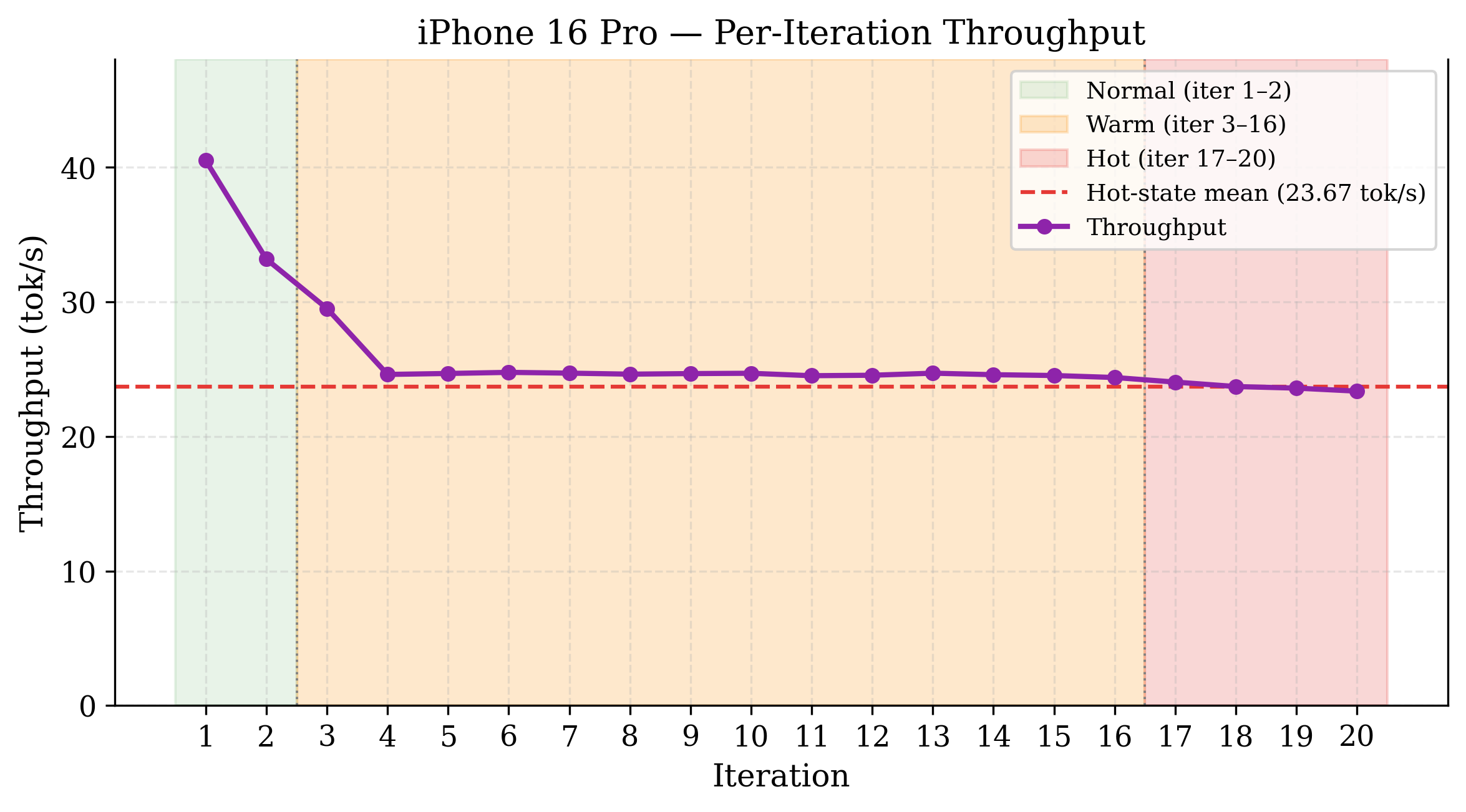}
\caption{iPhone~16 Pro per-iteration throughput across 20 iterations.
         Background shading indicates thermal state: green (Normal,
         iter 1--2), orange (Warm, iter 3--16), red (Hot, iter 17--20).
         Dashed line marks the sustained Hot-state mean of 23.67~tok/s.}
\label{fig:iphone_tps}
\end{figure}

The iPhone~16 Pro exhibits a clear three-phase thermal trajectory over 20
iterations. During the first two iterations (Normal state), throughput peaks
at 40.49~tok/s and averages 36.83~tok/s. By iteration~3 the device transitions
to a Warm state; across iterations 3--16 throughput averages 24.96~tok/s
($-$38.4\% from peak), stabilising within $\pm$1~tok/s after iteration~4.
From iteration~17 onwards the device enters the Hot state, where throughput
settles at $23.67 \pm 0.28$~tok/s (CV~= 1.2\%, $-$41.5\% from peak). The
overall throughput CV of 15.9\% is driven almost entirely by the
Normal-to-Warm transition; within the Warm and Hot plateaus CV drops below
2\%, comparable to the other platforms. The 1-second inter-iteration
cooldown is insufficient to permit any thermal recovery. Battery state of
charge drops 5\% over 20 iterations, projecting to approximately 400
inferences per full charge at this output length.

The observed thermal degradation is consistent with prior work on iOS LLM
inference. Independent benchmarking of Apple devices under sustained
inference~\cite{xiao2024llmpockets} finds that Apple GPU performance under
iOS exhibits noticeable fluctuations throughout the inference process, with
degradation occurring earlier and becoming more pronounced under long-prompt
workloads, consistent with our 258-token prompt driving a Normal-to-Warm
transition by iteration~3 and a full-Hot plateau by iteration~17. A separate
study on iPhone LLM inference~\cite{zhang2025cpugpu} explicitly places devices
in an ice-cooled environment to mitigate thermal throttling, noting that
real-world deployment conditions induce heat buildup and frequency scaling
not captured under controlled thermal conditions, corroborating our finding
that the 1-second inter-iteration gap is insufficient for thermal recovery
under natural operating conditions.

\FloatBarrier

\subsection{Samsung Galaxy S24 Ultra (Android / MLC-LLM)}

Table~\ref{tab:android_results} summarises results across 20 warm iterations
with greedy decoding, a 2{,}048-token context window, a 128-token prefill
chunk, and the display held off throughout measurement. The benchmark
completes all 20 iterations with no OS-enforced frequency floor event.

\begin{table}[H]
\centering
\caption{Samsung S24 Ultra inference performance (20 iterations, warm
         condition, greedy decoding, 2{,}048-token context, 128-token prefill
         chunk). Power and energy figures are collected via the Android
         \texttt{BatteryManager} API with the display off; absolute accuracy
         remains lower than hardware-instrumented measurement.}
\label{tab:android_results}
\begin{tabular}{lrrrr}
\toprule
\textbf{Metric} & \textbf{Mean} & \textbf{Std Dev} & \textbf{Min} & \textbf{Max} \\
\midrule
Decode tokens              & 646      & 0.3    & 645      & 646      \\
Decode time (ms)           & 59{,}908 & 4{,}122 & 52{,}906 & 67{,}619 \\
Throughput (tok/s)         & 10.83    & 0.75   & 9.55     & 12.21    \\
Prefill time (ms)          & 91{,}376 & 3{,}077 & 89{,}899 & 104{,}108 \\
Max GPU temperature (°C)   & 64.6     & 1.9    & 61.1     & 68.5     \\
Max CPU temperature (°C)   & 61.3     & 1.2    & 59.7     & 64.0     \\
GPU frequency (MHz)        & 804      & 75     & 720      & 1{,}000  \\
Avg power (mW)$^\ast$      & 1{,}590  & 198    & 1{,}222  & 2{,}075  \\
Energy per token (mJ)$^\ast$ & 146.4  & 10.8   & 127.9    & 175.1    \\
Battery drain (total)      & \multicolumn{4}{l}{7\% over 20 iterations (100\% $\to$ 93\%)} \\
\bottomrule
\end{tabular}
\end{table}

\noindent$^\ast$Derived from \texttt{BatteryManager} current/voltage with the
display off; see Section~\ref{sec:discussion} for measurement caveats.

\begin{figure}[H]
\centering
\includegraphics[width=0.72\textwidth]{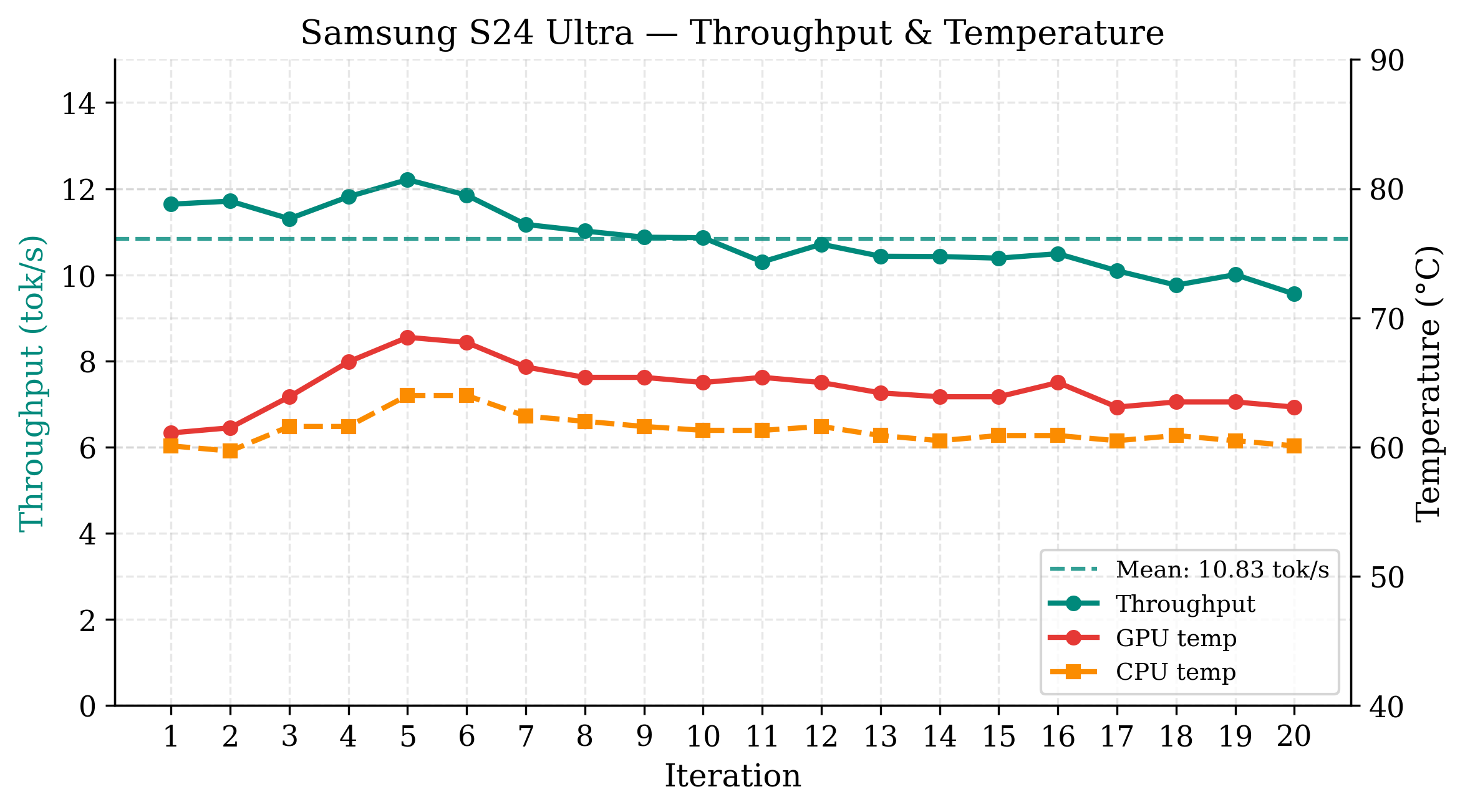}
\caption{Samsung S24 Ultra per-iteration throughput and temperature
         across all 20 iterations. Mean throughput of 10.83~tok/s
         (CV~= 6.9\%). No OS-enforced frequency floor observed in
         this run; GPU frequency varied between 720--1{,}000~MHz.}
\label{fig:s24_tps}
\end{figure}

Across the 20 iterations the S24 Ultra exhibits a gradual thermally-bounded
decline in throughput. Throughput peaks at 12.21~tok/s during a brief
Adreno DVFS boost (1{,}000~MHz, iterations~4--5) and settles into a
720--770~MHz plateau from iteration~8 onward, where throughput averages
$10.38 \pm 0.44$~tok/s (CV~= 4.2\%) with a minimum of 9.55~tok/s at
iteration~20---a 22\% degradation from peak. The full 20-iteration mean
is $10.83 \pm 0.75$~tok/s (CV~= 6.9\%).

Prefill time stabilises at $91{,}376 \pm 3{,}077$~ms (CV~= 3.4\%). This
figure is driven by the 128-token \texttt{prefill\_chunk\_size}, which
serialises OpenCL dispatch across many kernel invocations and is not a
reliable indicator of Adreno~750 hardware capability.

Thermal behaviour stays within the governor's sustained budget. Peak GPU
temperature reaches 68.5°C during the 1{,}000~MHz boost phase
(iteration~5) and subsequently stabilises at 63--65°C with the plateau
frequency. Peak CPU temperature follows a similar trajectory (59.7--64.0°C)
and neither trends upward after iteration~8. The governor regulates via
continuous DVFS rather than threshold-triggered shutdown, producing the
observed graceful degradation contour.

Average power sits at $1.59 \pm 0.20$~W with a clear downward trend
(1.93~W at iteration~1 to 1.22~W at iteration~20), reflecting the
frequency plateau and a cooler sustained GPU. Energy per token averages
$146 \pm 11$~mJ across 20 iterations, placing the S24 Ultra within the
same energy band as the RTX~4050 and Hailo-10H (297 and 271~mJ/token
respectively at the whole-system level), though the three measurements
remain methodologically distinct.

The observed throughput is broadly consistent with prior work on
Snapdragon~8~Gen~3 inference. Transformer-Lite~\cite{transformerlite2024}
reports decode throughput of 10--20~tok/s on Snapdragon~8~Gen~3 for
Qwen~1.5~4B under MLC-LLM, placing our 10.83~tok/s mean for the smaller 1.5B
model within the expected range and consistent with MLC-LLM's OpenCL
backend efficiency on Adreno. Prior work on mobile LLM
benchmarking~\cite{xiao2024llmpockets} specifically notes that MLC-LLM on
Adreno GPUs performs worse in prefill than llama.cpp on CPU, attributed to
the difficulty of optimising LLM operators across heterogeneous mobile GPU
architectures; we similarly attribute our inflated prefill figure to framework effects,
amplified here by the reduced chunk size, rather than to raw Adreno~750
capability. Independent mobile benchmarking
studies~\cite{xiao2024llmpockets} limit inference to five consecutive runs
per session with mandatory reboots between rounds to avoid overheating,
so achieving 20 consecutive iterations within the Android governor's
sustained budget required the display-off, bounded-context configuration
used here.

\FloatBarrier

\subsection{Cross-Platform Comparison}

Table~\ref{tab:crossplatform} summarises sustained inference performance
across all four platforms, and compares power
efficiency for the two platforms with reliable power measurements.

\begin{table}[H]
\centering
\caption{Cross-platform sustained inference performance and power
         efficiency summary. iPhone figures reflect sustained Hot state
         (iterations 17--20); S24 Ultra figures reflect the throttled
         plateau (iterations 8--20); RTX and Hailo figures reflect runs
         2--20 after cold-start exclusion. $^\dagger$~=~battery-throttled;
         power reflects GPU-level draw via \texttt{nvidia-smi}.
         $^\ddagger$~=~whole-system draw via INA219.
         $^\S$~=~\texttt{BatteryManager} with display off; absolute
         accuracy caveated (Section~\ref{sec:discussion}).
         $^*$~=~not available due to platform measurement limitations.}
\label{tab:crossplatform}
\small
\begin{tabularx}{\textwidth}{lXXXXX}
\toprule
\textbf{Platform} & \textbf{TPS} & \textbf{CV} & \textbf{Avg Power (W)} & \textbf{W/tok/s} & \textbf{mJ/token} \\
\midrule
RTX~4050$^\dagger$        & $131.70 \pm 2.87$ & 2.2\%  & 34.12$^\dagger$ & 0.259$^\dagger$ & 297.3$^\dagger$ \\
iPhone~16 Pro (Hot)       & $23.67 \pm 0.28$  & 1.2\%  & $^*$            & $^*$            & $^*$            \\
S24 Ultra (plateau)       & $10.38 \pm 0.44$  & 4.2\%  & 1.486$^\S$      & 0.143$^\S$      & 143.0$^\S$      \\
RPi~5 + Hailo-10H         & $6.914 \pm 0.003$ & 0.04\% & 1.870$^\ddagger$ & 0.271$^\ddagger$ & 270.5$^\ddagger$ \\
\midrule
\multicolumn{6}{l}{\small $^*$Power not available: iOS exposes no per-component API.} \\
\multicolumn{6}{l}{\small Energy figures are not directly comparable across platforms due to differing} \\
\multicolumn{6}{l}{\small measurement scope (GPU-level, whole-system, and display-off fuel-gauge).} \\
\bottomrule
\end{tabularx}
\end{table}

The RTX~4050 leads by a substantial margin at 131.7~tok/s, delivering
$19.0\times$ the throughput of the Hailo-10H, $5.6\times$ that of the
iPhone~16 Pro in its sustained Hot state, and $12.7\times$ that of the
S24 Ultra in its throttled plateau. Among mobile platforms, the
iPhone~16 Pro outperforms the S24 Ultra by $2.3\times$ in sustained
throughput.

The Hailo-10H achieves a throughput CV of 0.04\%, two orders of magnitude
lower than any other platform, a consequence of deterministic NPU scheduling
and fixed-length greedy decoding. The iPhone~16 Pro and S24 Ultra both reach
throughput CVs under 5\% once their respective thermal plateaus are entered,
though the \emph{path to} that plateau involves a 41.5\% drop on the iPhone and
a 15.0\% drop on the S24 Ultra relative to their own peak iterations. Among
platforms with reliable power measurements, the Hailo-10H and RTX~4050
exhibit near-identical energy proportionality: 0.271 and 0.259~W per tok/s
respectively. Despite an $18.2\times$ difference in absolute power draw and
a $19\times$ difference in throughput, the two platforms deliver equivalent
computation per joule, with energy per token of 270.5~mJ (Hailo-10H) versus
297.3~mJ (RTX~4050). The S24 Ultra's fuel-gauge-derived figure of 143~mJ/token
appears lower than both but is measured with the display off via
\texttt{BatteryManager}, which excludes panel draw and has coarser absolute
accuracy than hardware instrumentation; we report it as suggestive rather
than as a definitive comparison.

% ============================================================
\section{Discussion}
\label{sec:discussion}
% ============================================================

\subsection{Thermal Behaviour Under Sustained Load}

Sustained mobile LLM inference under our test protocol produces two
qualitatively different thermal outcomes rather than a single failure mode.
The iPhone~16 Pro demonstrates excellent peak throughput (40.49~tok/s)
but cannot sustain it: throughput drops $\sim$27\% within three iterations
as the device transitions to the Warm state, declines further over the
next thirteen, and settles at 23.67~tok/s in the Hot state for the final
four iterations, a 41.5\% reduction from peak. This is consistent with the
passive-cooling design of smartphones, where junction-to-ambient thermal
resistance is high and heat dissipation is constrained by form factor.

The S24 Ultra exhibits a milder curve regulated by the Android thermal
governor rather than a hard thermal failure. GPU frequency steps down
from a brief 1{,}000~MHz boost at iterations~4--5 to a 720--770~MHz
plateau by iteration~8, producing a 15\% throughput reduction from peak.
This is closer to the RTX~4050's own gradual decline from run~2 to run~20
than to the iPhone's steep Normal-to-Hot transition, and it is achieved
under a constrained configuration (display off, 2{,}048-token context,
128-token prefill chunk) chosen specifically to keep the device within
the governor's sustained budget. Under less constrained settings, earlier
runs hit the OS-enforced GPU frequency floor documented in the
methodology; the present result therefore shows that mobile sustained
inference is achievable but requires framework-level configuration
discipline.

Taken together, these results suggest thermal management is a primary
constraint for mobile LLM inference but not a uniform one. Peak
throughput is not sustainable on either smartphone in our test, yet the
practical degradation ranges from mild (S24 Ultra, 15\%) to severe
(iPhone, 41.5\%), and the operative ceiling for an always-on workload
is the plateau throughput rather than the peak. Both mobile platforms
share a one-second inter-iteration gap that does not permit thermal
recovery, so always-on agents querying more frequently than several
minutes apart would face either the plateau (S24) or the Hot-state
ceiling (iPhone) as the operational performance level.

\subsection{The Case for Dedicated Edge NPUs}

The Hailo-10H occupies a different point in the design space. At
6.914~tok/s and 1.87~W, it operates at $19\times$ lower throughput than
the battery-throttled RTX~4050 and 3.4$\times$ lower than the iPhone~16
Pro's Hot-state plateau, but it does so with near-zero performance
variance (CV~= 0.04\%), deterministic prefill time, no thermal
throttling, and a power envelope compatible with battery or low-wattage
PSU operation indefinitely. The cold-start model loading cost (11.86~s)
is a one-time overhead that does not affect steady-state operation.

Energy per token on the Hailo-10H (270.5~mJ, whole-system) is of the
same order as the RTX~4050's GPU-level figure (297.3~mJ) and the S24
Ultra's display-off fuel-gauge estimate (146~mJ), but we stress that
these three measurements are not directly comparable in scope
(Section~\ref{sec:limitations}). The defensible claim is that the
Hailo-10H does not trade off significantly worse energy-per-token for
its low power envelope, not that it achieves GPU-class per-joule
efficiency.

The principal limitation is latency. At 6.914~tok/s, a 500-token reply
takes $\sim$72 seconds. This is acceptable for background summarisation,
scheduled digest generation, or asynchronous agent tasks where the
user is not waiting on the response, but prohibitive for interactive
turn-by-turn dialogue. The 6.914~tok/s figure is also substantially
attributable to the deployment configuration rather than the NPU
itself, and gains are plausible through batched or speculative decoding
strategies adapted for NPU dispatch. The present result should therefore
be read as a conservative lower bound on Hailo-class NPU inference
rather than a ceiling; the thermal and efficiency properties
demonstrated here are likely to persist as throughput improves with
better host integration.

\subsection{Deployment Implications}

Table~\ref{tab:deployment} maps our empirical findings to deployment
scenarios under the tested framework and hardware configurations. The
four platforms occupy distinct points in the
throughput/power/availability space, and no single device dominates
across all scenarios.

The RTX~4050 delivers strong throughput and thermal stability where
AC power is available, but its 34~W sustained draw imposes meaningful
battery constraints. At 12\% battery drain over 20 inference runs,
continuous battery-powered operation would deplete a typical laptop
battery in approximately 2--3 hours under sustained workloads, and
genuine always-on battery deployment is not viable at this power
envelope. It is therefore rated marginal ($\sim$) for battery-powered
always-on use despite its performance headroom.

The iPhone~16 Pro is suitable for intermittent queries and sustains
its Hot-state plateau of 23.7~tok/s reliably once thermal equilibrium
is reached, but peak throughput is unsustainable and users relying on
burst performance will encounter a 41.5\% drop under sustained load.
The S24 Ultra completes all 20 iterations and settles at a $\sim$10~tok/s
plateau under the display-off, 2{,}048-token context, 128-token prefill
chunk configuration used here, demonstrating that mobile sustained
inference is workable with appropriate framework tuning. Deployments
that cannot adopt these constraints (for example, apps requiring the
display active) would face the earlier throttling behaviour rather
than the plateau.

The Hailo-10H offers a complementary rather than competitive operating
point. It is the only platform in our study that satisfies thermal
stability, low power draw, and consistent availability simultaneously
under its tested configuration, and at under 2~W it can sustain
continuous operation on a small battery pack or low-wattage PSU
indefinitely. Its trade-off is absolute throughput: at 6.914~tok/s it
is unsuitable for low-latency streaming responses but well-suited to
asynchronous or background inference, where consistent availability
matters more than response speed. For deployments that need
interactive latency \emph{and} always-on operation, none of the four
platforms in isolation is adequate; combinations of an NPU for
background work and a mobile SoC for interactive bursts may be a more
realistic architecture than any single device.

\begin{table}[H]
\centering
\caption{Deployment scenario suitability based on empirical results.
         Ratings reflect observed platform behaviour under tested
         framework configurations only. \checkmark~= suitable based on
         observed thermal stability and throughput; $\sim$~= marginal,
         either due to thermal degradation, low throughput, or framework
         limitations; \texttimes~= unsuitable based on observed thermal
         failure or throughput floor. S24 Ultra ratings assume the
         display-off, 2{,}048-token context, 128-token prefill chunk
         configuration documented here.}
\label{tab:deployment}
\small
\begin{tabular}{lcccc}
\toprule
\textbf{Scenario} & \textbf{RTX~4050} & \textbf{iPhone~16 Pro} & \textbf{S24 Ultra} & \textbf{RPi + Hailo} \\
\midrule
Interactive assistant (AC-powered)     & \checkmark & $\sim$     & $\sim$      & $\sim$     \\
Intermittent queries (5--10/hr)        & $\sim$     & \checkmark & \checkmark  & \checkmark \\
Sustained agent ($>$20/hr)             & $\sim$     & \texttimes & $\sim$      & \checkmark \\
Battery-powered always-on              & $\sim$     & \texttimes & $\sim$      & \checkmark \\
Low-latency streaming response         & \checkmark & $\sim$     & \texttimes  & \texttimes \\
\bottomrule
\end{tabular}
\end{table}

\subsection{Limitations and Threats to Validity}
\label{sec:limitations}

\paragraph{Framework heterogeneity.}
Each platform uses a different inference framework (vLLM, MLC-LLM,
MLX, hailo-ollama) and a different quantisation format (GPTQ~Int4,
q4f16\_2, MLX~Q4\_0, GGUF~Q4\_0). While all formats are nominally
4-bit, differences in grouping and calibration may introduce
confounding variation in output quality and throughput. More
fundamentally, observed performance differences reflect the combined
effect of hardware \emph{and} software rather than hardware alone.
The S24 Ultra's inflated prefill time, for example, is largely
framework-driven, and the RTX~4050's determinism is partly a
consequence of vLLM's scheduling behaviour rather than of the GPU
itself. Results should therefore be read as end-to-end platform
characterisations of what a practitioner would actually deploy, not
as hardware benchmarks in isolation. This heterogeneity mirrors
real-world deployment constraints, where the available framework is
largely fixed by the platform, and in that sense constitutes an
ecological validity strength rather than a pure weakness.

\paragraph{Single model, single prompt.}
The study evaluates one model (Qwen~2.5~1.5B) with one long-form
generation prompt. Shorter outputs, smaller or larger models, or
conversational prompts may exhibit different thermal trajectories
and plateau behaviour. Token generation lengths also differ across
platforms under greedy decoding (564--1{,}789 tokens), which affects
thermal load duration per iteration and complicates direct comparison
of thermal behaviour between devices even though the decoding policy
is identical.

\paragraph{S24 Ultra configuration constraints.}
The Android figures reflect a configuration adopted specifically to
keep the device within the thermal governor's sustained budget
across 20 iterations: the context window is set to 2{,}048 tokens,
the MLC-LLM prefill chunk is set to 128 tokens, and the display is
held off throughout the run. The reduced chunk size distributes
prefill compute across multiple smaller OpenCL dispatches rather than
one large parallel operation, lowering the peak GPU thermal spike
per iteration and avoiding the OS-enforced GPU frequency floor
observed in earlier unconstrained runs. This trade-off inflates the
reported prefill time well beyond the Adreno~750's native compute
capability, and the reduced context window bounds the KV cache more
tightly than on the other platforms. The S24 Ultra's sustained
plateau should therefore be read as achievable under deliberate
framework tuning, not as the device's behaviour under default
app-developer settings; deployments that cannot adopt these
constraints would likely encounter the earlier throttling behaviour.

\paragraph{Power measurement methodology.}
Power measurement is not uniform across platforms, which limits
cross-platform energy comparison. The RTX~4050 uses \texttt{nvidia-smi}
GPU power reporting at 100~ms intervals, capturing GPU-level draw with
reasonable accuracy. The RPi~5 + Hailo-10H uses an INA219 current
sensor on the PMIC supply rails, reflecting whole-system draw rather
than isolated accelerator power. The S24 Ultra uses the Android
\texttt{BatteryManager} fuel gauge with the display held off; values
trend predictably with GPU frequency and temperature but absolute
accuracy is lower than hardware-instrumented measurement, and the
figure excludes panel draw and any leakage recorded against other
system domains. The iPhone~16 Pro exposes no per-component power
API; battery state-of-charge serves as a coarse energy proxy. As a
result, the energy-per-token figures reported for the three
power-instrumented platforms are not directly comparable: the RTX
figure reflects GPU-level power, the S24 figure reflects fuel-gauge
battery draw with display off, and the Hailo figure reflects
whole-system power including the RPi~5 SoC. True accelerator-level
efficiency on each platform is likely different from the reported
figures, and the apparent energy-per-token similarities between
platforms should be read as suggestive rather than definitive.
Standardised power instrumentation across edge platforms remains an
open methodological challenge.

\paragraph{Other threats to validity.}
The RTX~4050 was benchmarked on battery, so its reported throughput
and power reflect battery-throttled performance rather than the GPU's
full 75~W TGP capability. Cold-start behaviour was recorded but not
analysed; only steady-state results are reported. Only one unit per
platform was tested, so device-to-device variability in thermal
performance, binning, and battery condition is not characterised. The
20-iteration session length is sufficient to observe plateau
behaviour on all platforms but not to characterise plateau drift over
longer horizons. Finally, all observations reflect platform-level
behaviour under specific inference frameworks and cannot be
attributed to hardware in isolation; alternative software stacks may
yield materially different results on the same devices.

\subsection{Conclusion and Future Work}

Sustained LLM inference on edge platforms is constrained by thermal
behaviour, power, and memory bandwidth in different proportions across
the four devices tested. The iPhone~16 Pro loses 41.5\% of its peak
throughput over 20 iterations and settles into a Hot-state plateau
at 23.7~tok/s. The S24 Ultra completes 20 iterations with a milder
15\% gradual DVFS-regulated decline to a $\sim$10~tok/s plateau under
our constrained configuration, demonstrating that mobile sustained
inference is achievable when framework-level tuning respects the
governor's sustained budget. The RTX~4050 sustains 131.7~tok/s at
battery-throttled 34~W but cannot meaningfully operate as an
always-on device on battery. The Hailo-10H sustains 6.914~tok/s at
under 2~W with effectively zero variance, providing thermally stable
asynchronous inference at the cost of interactive latency.

These results do not support a single ``best'' edge platform
recommendation. Rather, the four devices occupy distinct points in
the throughput/power/availability space, and matching a platform to a
use case requires explicitly characterising whether peak throughput,
sustained throughput, battery life, or response latency is the
binding requirement.

Future work will characterise plateau stability over 100+ iterations
to test whether the observed plateaus drift over longer horizons,
replace the Android Battery Manager API with a hardware current sensor
for more precise power measurement, investigate duty-cycling and
active cooling strategies for mobile deployment, evaluate the impact
of quantisation format unification across platforms, and assess
additional models and prompt types to broaden generalisability.

% ============================================================

\end{document}